# High precision indoor positioning by means of LiDAR

**<u>E. Sánchez[1]</u>, M. Botsch[1], B. Huber [2], A. García[3]**

[1] Technische Hochschule Ingolstadt
Esplanade 10
85049 Ingolstadt
GERMANY

[2] GeneSys Elektronik GmbH
In der Spöck 10
77656 Offenburg
GERMANY

[3] Autolog Group, Universidad de Castilla-La Mancha
C/ Altagracia, 50
13071 Ciudad Real
SPAIN

**Abstract**

The trend towards autonomous driving and the continuous research in the automotive area, like Advanced Driver Assistance Systems (ADAS), requires an accurate localization under all circumstances. An accurate estimation of the vehicle state is a basic requirement for any trajectory-planning algorithm. Still, even when the introduction of the GPS L5 band promises lane-accuracy, coverage limitations in roofed areas still have to be addressed.

In this work, a method for high precision indoor positioning using a LiDAR is presented. The method is based on the combination of motion models with LiDAR measurements, and uses infrastructural elements as positioning references. This allows to estimate the orientation, velocity over ground and position of a vehicle in a Local Tangent Plane (LTP) reference frame. When the outputs of the proposed method are compared to those of an Automotive Dynamic Motion Analyzer (ADMA), mean errors of 1°, 0.1 m/s and of 4.7 cm respectively are obtained. The method can be implemented by using a LiDAR sensor as a stand-alone unit. A median runtime of 40.77 µs on an Intel i7-6820HQ CPU signals the possibility of real-time processing.

## 1. Introduction

Important trends of the automotive industry are ADAS and autonomous driving. For both cases, certain tasks are already implemented in production vehicles, such as lane keeping assist or autonomous emergency braking. However, there is continuous research interest on driving functions, such as autonomous parking. Considering that the vehicle state estimation is a primary requirement for vehicle trajectory-planning, there is a need for accurate positioning systems.

Nonetheless, even when the introduction of the GPS L5 band and other highly accurate satellite positioning systems promise lane-accuracy, the coverage limitations are still present. Roofed areas (such as tunnels or multi-story parking lots), urban and natural canyons are still weak points of satellite navigation. This creates a need for accurate Indoor Positioning Systems (IPSs).

Another motivation for using IPSs is the need of creating artificial environments for sensors or vehicle-functions testing. For example, one might want to create a reproducible environment that affects the vehicle sensors, in order to test their performance.

Yet another highly relevant application of high accuracy IPSs is to serve as a reference for evaluating other navigation or positioning systems. As stated above, current SatNav reference technology is not available in roofed areas, such as enclosed test facilities. A highly accurate IPS has the potential of substituting SatNav reference systems for indoor applications.

In this work, a theoretical and experimental approach is taken for creating a method for an IPS that has a high accuracy. The accuracy is computed by comparing the outputs of the proposed method with those of an ADMA-G-Pro+. This Inertial Navigation System (INS) integrates servo-accelerometers and optical gyroscopes, and gets correction data from a SatNav receiver with Real-Time Kinematic (RTK).

This paper makes the following contributions: It shows 1) the detailed mathematical procedure for projecting the measurements of an Inertial Measurement Unit (IMU) on the Local Tangent Plane (LTP), 2) the detailed mathematical derivation of a novel, robust, highly accurate LiDAR-based Positioning Method (LbPM), 3) a benchmark of the proposed method against an ADMA-G-Pro+, 4) indoor testing of the proposed method, and 4) runtime measurements of the proposed method.

The paper is structured in the following manner: first a brief overview of the state-of-the-art of indoor positioning is given. Then, the mathematical derivation of the proposed algorithm is shown. Afterwards, an accuracy benchmark of the proposed method is made and indoor testing is shown. Finally, conclusions and future work are addressed.

## 2. State-of-the-art

As explained above, indoor positioning is a topic of interest because of the limitations of existing SatNav reference systems as well as the need for research in enclosed areas. Some of the latest research in the field can be found in [1] and [2]. The methods presented in these publications combine a series of wireless communication protocols, proprietary technologies like iBeacon [3] and other hardware for position estimation.

The LiDAR positioning method with the highest accuracy found is presented in [4]. This method is thought specifically for vehicle position and velocity estimation. Its functioning principle is to place a number of LiDARs at ground level so as to avoid occlusions with the same vehicles. The points seen by the LiDAR are then classified as either *active* or *static*. Static points are removed, as they represent walls, pillars and other fixed infrastructure elements. From the resulting active points, the wheels are recognized by shape matching. Later, assuming vehicles with four wheels, a pose (position and orientation) can be estimated. An accuracy of 11.5 cm, with a standard deviation $(\sigma)$ of 5.4 cm is shown. The outputs of the algorithm where compared to "human-labelled" ground truth.

The indoor positioning method with the highest accuracy found is the Active Bat [5,6]. The system is based on ultra-sonic waves, and has two main elements: an emitter and an array of receivers. The emitter is a spherical-shaped ultra-sonic emitter that is oriented so it emits the ultrasonic waves upwards. An array of receivers are then placed on the roof of the area where the emitters are to be located. By using the Time of Flight (ToF) and trilateration, the receiver is located. An accuracy with a mean error of 3 cm is shown, but the evaluation methodology is not explained in detail [7]. One of the biggest disadvantages of the Active Bat, is the number of receivers required. They have to be located on a square pattern with separations of 1.2 m between receivers. Considering the typical use-cases of the automotive industry, the sheer number of receivers required can make this approach impractical.

## 3. Mathematical preamble

The coordinate systems used in this work are the Local Tangent Plane and the Local Vehicle Plane (LCP). These are defined in the following. The LTP is defined similar to the East-North-Up (ENU) reference frame. It is composed by the $x_{\mathrm{LTP}}$, $y_{\mathrm{LTP}}$ and $z_{\mathrm{LTP}}$ axes, pointing East, North and Up respectively; with $\overrightarrow{z_{\mathrm{LTP}}} = \overrightarrow{x_{\mathrm{LTP}}} \times \overrightarrow{y_{\mathrm{LTP}}}$, and arbitrary origin $o_{\mathrm{LTP}}$ on the surface of the Earth. The Local Vehicle Plane is defined similar to the ISO8855:2011 norm. It is composed by the $x_{\mathrm{LCP}}$, $y_{\mathrm{LCP}}$ and $z_{\mathrm{LCP}}$ axes, pointing towards the hood, the driver side and Up respectively; with $\overrightarrow{z_{\mathrm{LCP}}} = \overrightarrow{x_{\mathrm{LCP}}} \times \overrightarrow{y_{\mathrm{LCP}}}$, and origin $o_{\mathrm{LCP}}$ on the Center of sprung Mass (CoM) of the vehicle. A graphical representation of the LCP can be seen in the Figure 1. For simplification purposes, it is assumed that all sensors mounted on the vehicle use the same reference frame as the LCP.

Vectors are represented in boldface and matrices in boldface, capital letters. All the units are given in the International System of Units (SI), unless otherwise specified.

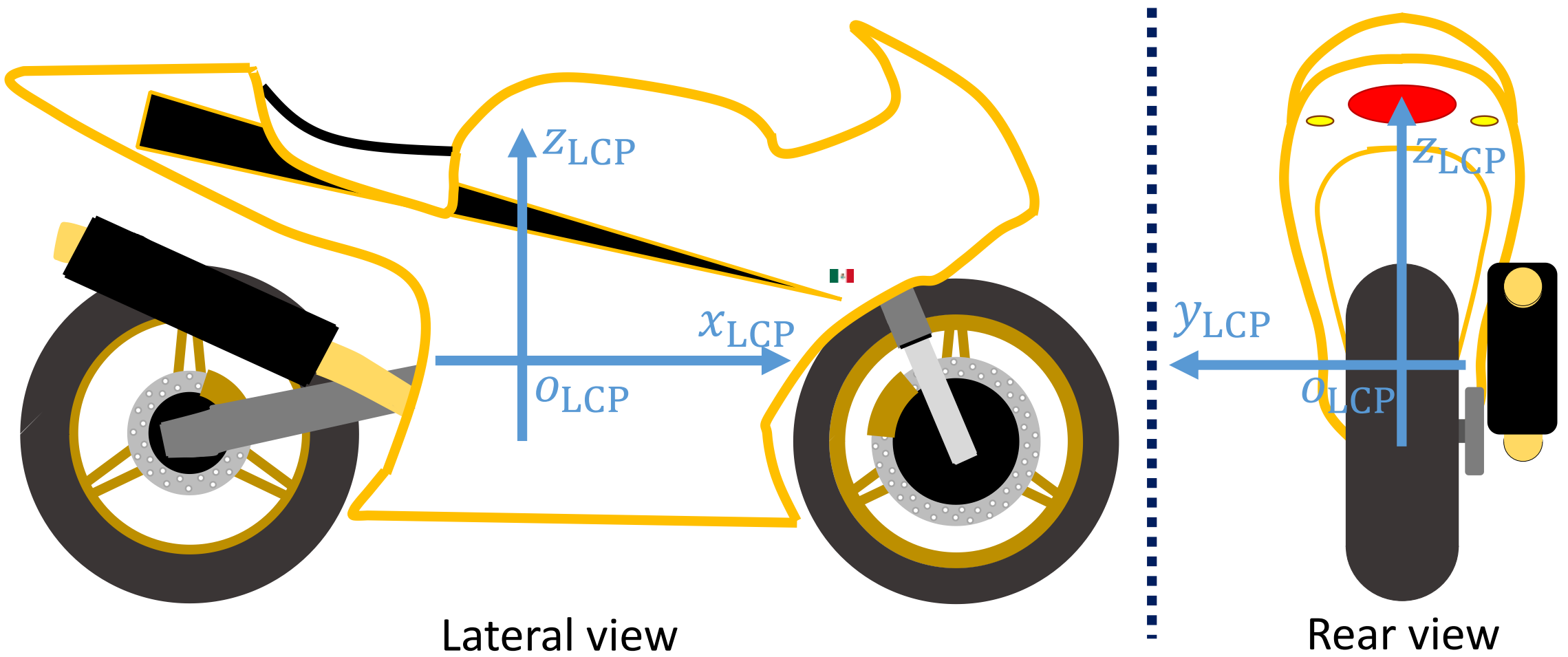

Figure 1. LCP coordinate frame.

## 4. Horizontation of inertial measurements

For this work, horizontation is understood as the projection of the sensor measurements as if the $(x_{\mathrm{LCP}}, y_{\mathrm{LCP}})$–plane was parallel to the $(x_{\mathrm{LTP}}, y_{\mathrm{LTP}})$–plane, and the $z_{\mathrm{LCP}}$ axis had the same orientation as the $z_{\mathrm{LTP}}$ axis. This is required because the motion models used express the vehicle movement as if it was planar on the $(x_{\mathrm{LTP}}, y_{\mathrm{LTP}})$–plane. Also, it is impractical to physically orient the sensors to the LTP, and to keep them with this orientation for the whole measurement duration. The horizontation procedure consists mainly on tracking the LTP axes with respect to the LCP, and is done by means of IMUs, but is applied to all sensors. The detailed procedure is explained in the following.

### *4.1. Horizontation at standstill*

The horizontation is initialized when the vehicles are at standstill. The reason for this is that the gravity vector can be used as reference. In the context of this work, standstill is defined as null velocity, acceleration and rotation rate with respect to the LTP. During initialization, all live loads should be avoided and all dead loads are irrelevant. Under these conditions, the only quantity measured by the accelerometers of the IMU is the gravity of the Earth, which magnitude is given by

$$G = \sqrt{a_{x_{\mathrm{LCP}}}^2 + a_{y_{\mathrm{LCP}}}^2 + a_{z_{\mathrm{LCP}}}^2}\,, \tag{1}$$

where $a_{x_{\mathrm{LCP}}}, a_{y_{\mathrm{LCP}}},$ and $a_{z_{\mathrm{LCP}}}$ are the instantaneous accelerations measured by the IMU along its axes. It is known to the authors that $G \approx 9.8\,\frac{m}{s^2}$, and that it varies slightly according to $o_{\mathrm{LTP}}$ because of the height (distance to the center of mass of the Earth), its position (relative location to the rotation axis of the Earth) and other local anomalies. On the other hand, high resolution gravity maps are available in the literature, should they be required to expand the methods here presented [8]. The first rotation to perform, is to align the $z_{\mathrm{LCP}}$ axis with the $z_{\mathrm{LTP}}$ axis. This can be expressed as a rotation around an axis, as shown in Figure 2. For this, the required rotation angle $\theta_{\mathrm{inc}}$ is calculated using a direction cosine as follows

$$\theta_{\mathrm{inc}} = \arccos\left(\frac{a_{z_{\mathrm{LCP}}}}{G}\right). \tag{2}$$

The required rotation axis $\boldsymbol{r}_{\mathrm{inc}_{\mathrm{LCP}}}$ is given by the vector

$$\boldsymbol{r}_{\mathrm{inc}_{\mathrm{LCP}}} = [a_{x_{\mathrm{LCP}}} \quad a_{y_{\mathrm{LCP}}} \quad a_{z_{\mathrm{LCP}}}] \times [0 \quad 0 \quad 1] = [r_x \quad r_y \quad r_z]. \tag{3}$$

The rotation matrix that rotates the LCP so $z_{\mathrm{LTP}}$ and $z_{\mathrm{LCP}}$ have the same orientation and direction is then given by

$$\boldsymbol{R}_{\mathrm{LCP}}^{\mathrm{LCP}'} = \begin{bmatrix} \frac{r_x^2+(r_y^2+r_z^2)\mathrm{c}(\theta_{\mathrm{inc}})}{r_{\mathrm{m}}} & \frac{r_x r_y r_{\mathrm{c}} - r_z\sqrt{r_{\mathrm{m}}}\mathrm{s}(\theta_{\mathrm{inc}})}{r_{\mathrm{m}}} & \frac{r_x r_z r_{\mathrm{c}} + r_y\sqrt{r_{\mathrm{m}}}\mathrm{s}(\theta_{\mathrm{inc}})}{r_{\mathrm{m}}} \\ \frac{r_x r_y r_{\mathrm{c}} + r_z\sqrt{r_{\mathrm{m}}}\mathrm{s}(\theta_{\mathrm{inc}})}{r_{\mathrm{m}}} & \frac{r_y^2+(r_x^2+r_z^2)\mathrm{c}(\theta_{\mathrm{inc}})}{r_{\mathrm{m}}} & \frac{r_y r_z r_{\mathrm{c}} - r_x\sqrt{r_{\mathrm{m}}}\mathrm{s}(\theta_{\mathrm{inc}})}{r_{\mathrm{m}}} \\ \frac{r_x r_z r_{\mathrm{c}} - r_y\sqrt{r_{\mathrm{m}}}\mathrm{s}(\theta_{\mathrm{inc}})}{r_{\mathrm{m}}} & \frac{r_y r_z r_{\mathrm{c}} + r_x\sqrt{r_{\mathrm{m}}}\mathrm{s}(\theta_{\mathrm{inc}})}{r_{\mathrm{m}}} & \frac{r_z^2+(r_x^2+r_y^2)\mathrm{c}(\theta_{\mathrm{inc}})}{r_{\mathrm{m}}} \end{bmatrix}, \text{ with} \tag{4}$$

$$r_{\mathrm{c}} = 1 - \mathrm{c}(\theta_{\mathrm{inc}}), \text{ and} \tag{5}$$

$$r_{\mathrm{m}} = r_x^2 + r_y^2 + r_z^2, \tag{6}$$

where $\mathrm{s}(\cdot)$ and $\mathrm{c}(\cdot)$ represent the sine and cosine functions. The step by step derivation of a matrix that rotates a vector around an arbitrary axis, also known as the Rodriguez' Formula, can be found in [9].

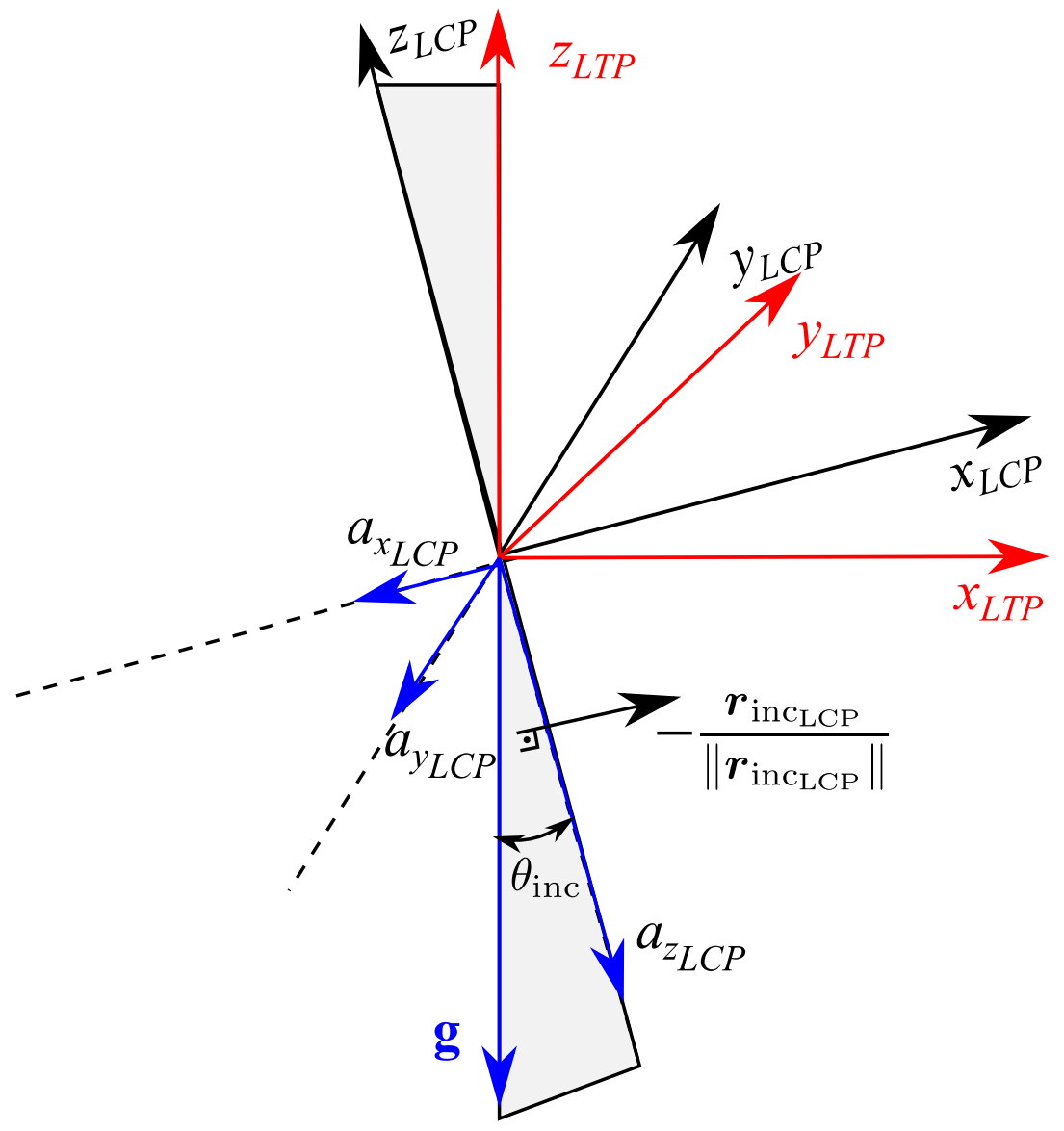

Figure 2. Shown is a rotation to align the $z_{\mathrm{LCP}}$ axis with the $z_{\mathrm{LTP}}$ axis.

The next orientation angle required is the yaw $\theta_{\mathrm{yaw}_{\mathrm{LCP}}}$. For this work, the yaw is defined as the angle between $x_{\mathrm{LTP}}$ and the projection on the ($x_{\mathrm{LTP}}$, $y_{\mathrm{LTP}}$)–plane of $x_{\mathrm{LCP}}$, measured from $x_{\mathrm{LTP}}$ and according to the right-hand rule. The quickest way to obtain an initial measurement is by reading it from another sensor that measures the Course Over Ground (COG), as can be a GPS receiver. The COG is defined in this work as the angle between the projection on the $(x_{\mathrm{LTP}}, y_{\mathrm{LTP}})$–plane of the velocity over ground $v_{o_{\mathrm{LCP}}}$ and $x_{\mathrm{LTP}}$, measured from $x_{\mathrm{LTP}}$ and according to the right-hand rule. Figure 3 is visualizing this angle. For this, the vehicle should be driven on a straight line so that $\theta_{\mathrm{yaw}_{\mathrm{LCP}}} = \mathrm{COG}$. This is applicable in use-cases where the GPS signal is available at the beginning of the measurements, as is the case for tunnels, or when driving into a closed parking house from the outside. Should this be not possible, an initial value for $\theta_{\mathrm{yaw}_{\mathrm{LCP}}}$ can be manually set. This value should have a maximum deviation of $\pm 45°$ from the real value, as this is the tolerance of the presented method. In the Figure 3, a drifting vehicle is depicted to show a clear difference between $\theta_{\mathrm{yaw}_{\mathrm{LCP}}}$ and the COG.

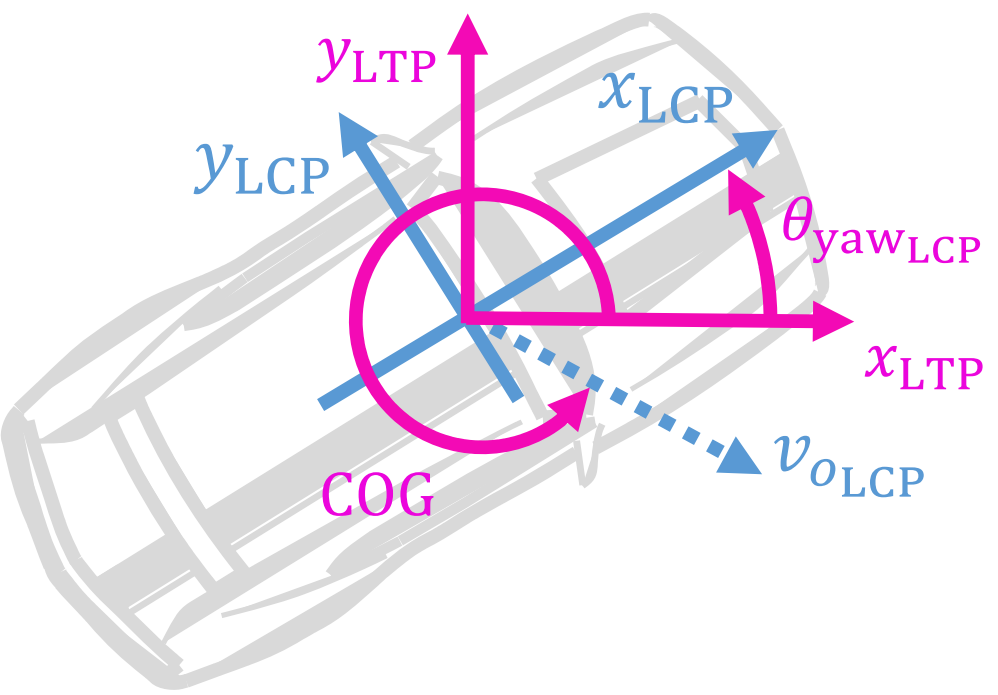

Figure 3. A drifting vehicle. Shown in blue is the LCP and the velocity over ground. Shown in pink is the LTP, the yaw angle and the Course Over Ground.

Knowing $\theta_{\mathrm{yaw}_{\mathrm{LCP}}}$, the initial value of the rotation matrix that tracks the LTP axes with respect to the LCP is given by

$$\boldsymbol{R}_{\mathrm{LTP}}^{\mathrm{LCP},\tau 0} = \left(\boldsymbol{R}\left(\theta_{\mathrm{yaw}_{\mathrm{LCP}}}\right) \boldsymbol{R}_{\mathrm{LCP}}^{\mathrm{LCP}'}\right)^{\mathrm{T}}, \text{ with} \quad (7)$$

$$\boldsymbol{R}\left(\theta_{\mathrm{yaw}_{\mathrm{LCP}}}\right) = \begin{bmatrix} \cos\left(\theta_{\mathrm{yaw}_{\mathrm{LCP}}}\right) & -\sin\left(\theta_{\mathrm{yaw}_{\mathrm{LCP}}}\right) & 0 \\ \sin\left(\theta_{\mathrm{yaw}_{\mathrm{LCP}}}\right) & \cos\left(\theta_{\mathrm{yaw}_{\mathrm{LCP}}}\right) & 0 \\ 0 & 0 & 1 \end{bmatrix}. \quad (8)$$

Then, considering the LTP axes as unity vectors, their positions in LCP at this initialization point are given by

$$\boldsymbol{x}_{\mathrm{LTP}}^{\mathrm{LCP},\tau 0} = \boldsymbol{R}_{\mathrm{LTP}}^{\mathrm{LCP},\tau 0}[1 \quad 0 \quad 0]^{\mathrm{T}}, \quad (9)$$

$$\boldsymbol{y}_{\mathrm{LTP}}^{\mathrm{LCP},\tau 0} = \boldsymbol{R}_{\mathrm{LTP}}^{\mathrm{LCP},\tau 0}[0 \quad 1 \quad 0]^{\mathrm{T}}, \text{ and} \quad (10)$$

$$\boldsymbol{z}_{\mathrm{LTP}}^{\mathrm{LCP},\tau 0} = \boldsymbol{R}_{\mathrm{LTP}}^{\mathrm{LCP},\tau 0}[0 \quad 0 \quad 1]^{\mathrm{T}}. \quad (11)$$

This procedure applies equally to all vehicles, regardless of a non-vertical initial position. This is the case, for example, for two-wheeled ground vehicles.

*4.2. Horizontation in motion*

Even when driving with constant velocity on a straight line, ground vehicles experiment random accelerations on all its axes. This can come from uneven roads, for example. This complicates the update of $\boldsymbol{R}_{\mathrm{LTP}}^{\mathrm{LCP},\tau 0}$ using the accelerometers. So, the update is done mainly using the gyroscopes. The process is detailed in the following.

With the measurements of the IMU at time instance $\tau 1$ being $\omega_{x_{\mathrm{LCP}}}$ , $\omega_{y_{\mathrm{LCP}}}$ and $\omega_{z_{\mathrm{LCP}}}$ , which are the instantaneous rotation rates around the $x_{\mathrm{LCP}}$, $y_{\mathrm{LCP}}$ and $z_{\mathrm{LCP}}$ axes, the skew symmetric matrix $\boldsymbol{S}\left(\boldsymbol{\omega}_{\mathrm{LCP},\tau 1}^{\mathrm{LCP},\tau 0}\right)$ of the composition of rotations of LCP is written as

$$\boldsymbol{S}\left(\boldsymbol{\omega}_{\mathrm{LCP},\tau 1}^{\mathrm{LCP},\tau 0}\right) = \begin{bmatrix} 0 & -\omega_{z_{\mathrm{LCP}}} & \omega_{y_{\mathrm{LCP}}} \\ \omega_{z_{\mathrm{LCP}}} & 0 & -\omega_{x_{\mathrm{LCP}}} \\ -\omega_{y_{\mathrm{LCP}}} & \omega_{x_{\mathrm{LCP}}} & 0 \end{bmatrix}, \text{ with} \tag{12}$$

$$\boldsymbol{\omega}_{LCP,\tau 1}^{LCP,\tau 0} = \begin{bmatrix} \omega_{x_{\mathrm{LCP}}} & \omega_{y_{\mathrm{LCP}}} & \omega_{z_{\mathrm{LCP}}} \end{bmatrix} . \tag{13}$$

The update of $\boldsymbol{R}_{\mathrm{LTP}}^{\mathrm{LCP},\tau 0}$, for the next time instance $\tau 1$ is then given by

$$\widehat{\boldsymbol{R}}_{\mathrm{LTP}}^{\mathrm{LCP},\tau 1} = \boldsymbol{R}_{\mathrm{LCP},\tau 0}^{\mathrm{LCP},\tau 1}\boldsymbol{R}_{\mathrm{LTP}}^{\mathrm{LCP},\tau 0} = \left(\boldsymbol{R}_{\mathrm{LCP},\tau 0}^{\mathrm{LCP},\tau 0} + \Delta\tau\, \dot{\boldsymbol{R}}_{\mathrm{LCP},\tau 0}^{\mathrm{LCP},\tau 1}\right)\boldsymbol{R}_{\mathrm{LTP}}^{\mathrm{LCP},\tau 0} , \tag{14}$$

$$\widehat{\boldsymbol{R}}_{\mathrm{LTP}}^{\mathrm{LCP},\tau 1} = \left(\boldsymbol{R}_{\mathrm{LCP},\tau 0}^{\mathrm{LCP},\tau 0} + \Delta\tau\, \boldsymbol{S}\left(\boldsymbol{\omega}_{\mathrm{LCP},\tau 0}^{\mathrm{LCP},\tau 1}\right)\boldsymbol{R}_{\mathrm{LCP},\tau 0}^{\mathrm{LCP},\tau 1}\right)\boldsymbol{R}_{\mathrm{LTP}}^{\mathrm{LCP},\tau 0} , \tag{15}$$

$$\widehat{\boldsymbol{R}}_{\mathrm{LTP}}^{\mathrm{LCP},\tau 1} = \left(\boldsymbol{R}_{\mathrm{LCP},\tau 0}^{\mathrm{LCP},\tau 0} + \Delta\tau\, \boldsymbol{S}\left(-\boldsymbol{\omega}_{\mathrm{LCP},\tau 1}^{\mathrm{LCP},\tau 0}\right)\boldsymbol{R}_{\mathrm{LCP},\tau 0}^{\mathrm{LCP},\tau 1}\right)\boldsymbol{R}_{\mathrm{LTP}}^{\mathrm{LCP},\tau 0} , \tag{16}$$

$$\widehat{\boldsymbol{R}}_{\mathrm{LTP}}^{\mathrm{LCP},\tau 1} = \left(\boldsymbol{I}_{3x3} + \Delta\tau\, \left(\boldsymbol{S}\left(\boldsymbol{\omega}_{\mathrm{LCP},\tau 1}^{\mathrm{LCP},\tau 0}\right)\right)^{T} \boldsymbol{I}_{3x3}\right)\boldsymbol{R}_{\mathrm{LTP}}^{\mathrm{LCP},\tau 0} , \tag{17}$$

$$\widehat{\boldsymbol{R}}_{\mathrm{LTP}}^{\mathrm{LCP},\tau 1} = \left(\Delta\tau \boldsymbol{S}\left(\boldsymbol{\omega}_{\mathrm{LCP},\tau 1}^{\mathrm{LCP},\tau 0}\right)^{\mathrm{T}} + \boldsymbol{I}_{3x3}\right)\boldsymbol{R}_{\mathrm{LTP}}^{\mathrm{LCP},\tau 0} , \tag{18}$$

where $\Delta\tau = \tau 1 - \tau 0$, $\boldsymbol{I}_{3x3}$ is the identity matrix, and it has been used that the time derivative of a rotation matrix $\boldsymbol{R}$ is $\dot{\boldsymbol{R}} = \boldsymbol{S}(\boldsymbol{\omega})\boldsymbol{R}$. Equation (18) can then be used for further updates according to $\boldsymbol{R}_{\mathrm{LTP}}^{\mathrm{LCP},\tau i} = \left(\Delta\tau \boldsymbol{S}\left(\boldsymbol{\omega}_{\mathrm{LCP},\tau \mathrm{i}}^{\mathrm{LCP},\tau(\mathrm{i}-1)}\right)^{\mathrm{T}} + \boldsymbol{I}_{3x3}\right)\boldsymbol{R}_{\mathrm{LTP}}^{\mathrm{LCP},\tau(i-1)}$.

In order to make the matrix $\widehat{\boldsymbol{R}}_{\mathrm{LTP}}^{\mathrm{LCP},\tau 1}$ orthonormal, i. e., an rotation matrix, the following steps are performed after each update

$$\widehat{\boldsymbol{R}}_{\mathrm{LTP}}^{\mathrm{LCP},\tau 1} = [\boldsymbol{r}_1{}' \quad \boldsymbol{r}_2{}' \quad \boldsymbol{r}_3{}'], \tag{19}$$

$$\boldsymbol{r}_3{}'' = \boldsymbol{r}_3{}', \tag{20}$$

$$\boldsymbol{r}_1{}'' = \boldsymbol{r}_2' \times \boldsymbol{r}_3{}'', \tag{21}$$

$$\boldsymbol{r}_2{}'' = \boldsymbol{r}_3'{}' \times \boldsymbol{r}_1{}'', \tag{22}$$

$$\boldsymbol{r}_1 = \frac{r_1''}{\|r_1''\|}, r_2 = \frac{r_2''}{\|r_2''\|}, \boldsymbol{r}_3 = \frac{r_3''}{\|r_3''\|}, \tag{23}$$

$$\boldsymbol{R}_{\mathrm{LTP}}^{\mathrm{LCP},\tau 1} = [\boldsymbol{r}_1 \quad \boldsymbol{r}_2 \quad \boldsymbol{r}_3]. \tag{24}$$

Next, the positions of the LTP axes with respect to the LCP are updated as follows

$$\boldsymbol{x}_{\mathrm{LTP}}^{\mathrm{LCP},\tau 1} = \boldsymbol{R}_{\mathrm{LTP}}^{\mathrm{LCP},\tau 1}[1 \quad 0 \quad 0]^{\mathrm{T}}, \tag{25}$$

$$\boldsymbol{y}_{\mathrm{LTP}}^{\mathrm{LCP},\tau 1} = \boldsymbol{R}_{\mathrm{LTP}}^{\mathrm{LCP},\tau 1}[0 \quad 1 \quad 0]^{\mathrm{T}}, \text{ and} \tag{26}$$

$$\boldsymbol{z}_{\mathrm{LTP}}^{\mathrm{LCP},\tau 1} = \boldsymbol{R}_{\mathrm{LTP}}^{\mathrm{LCP},\tau 1}[0 \quad 0 \quad 1]^{\mathrm{T}}. \tag{27}$$

The acceleration vector $\boldsymbol{a}_{\mathrm{LTP},\tau 1}$ and the angular velocity vector $\dot{\boldsymbol{\theta}}_{\mathrm{LTP},\tau 1}$ in LTP at time instance $\tau 1$ are given by

$$\boldsymbol{a}_{\mathrm{LTP},\tau 1} = [a_{x_{\mathrm{LTP},\tau 1}} \quad a_{y_{\mathrm{LTP},\tau 1}} \quad a_{z_{\mathrm{LTP},\tau 1}}], \text{ and} \tag{28}$$

$$\dot{\boldsymbol{\theta}}_{\mathrm{LTP},\tau 1} = \left[\dot{\theta}_{x_{\mathrm{LTP},\tau 1}} \quad \dot{\theta}_{y_{\mathrm{LTP},\tau 1}} \quad \dot{\theta}_{z_{\mathrm{LTP},\tau 1}}\right], \text{ where} \tag{29}$$

$$\boldsymbol{a}_{\mathrm{LTP},\tau 1} = \left(\boldsymbol{R}_{\mathrm{LTP}}^{\mathrm{LCP},\tau 1}\right)^{\mathrm{T}} \boldsymbol{a}_{\mathrm{LCP}}, \tag{30}$$

$$\dot{\boldsymbol{\theta}}_{\mathrm{LTP},\tau 1} = \left(\boldsymbol{R}_{\mathrm{LTP}}^{\mathrm{LCP},\tau 1}\right)^{\mathrm{T}} \boldsymbol{\omega}_{\mathrm{LCP},\tau 1}^{\mathrm{LCP},\tau 0}, \text{ and} \tag{31}$$

$$\boldsymbol{a}_{\mathrm{LCP}} = \left[a_{x_{\mathrm{LCP}}} \quad a_{y_{\mathrm{LCP}}} \quad a_{z_{\mathrm{LCP}}}\right]^{\mathrm{T}}. \tag{32}$$

Then, the magnitude $a_{\mathrm{OG}}$ and the orientation $\theta_{a_{\mathrm{OG}}}$ of the projected acceleration on the LTP are obtained as

$$a_{\mathrm{OG}} = \sqrt{a_{x_{\mathrm{LTP},\tau 1}}^2 + a_{y_{\mathrm{LTP},\tau 1}}^2}, \text{ and} \tag{33}$$

$$\theta_{a_{\mathrm{OG}}} = \operatorname{atan2}\left(\frac{a_{y_{\mathrm{LTP},\tau 1}}}{a_{x_{\mathrm{LTP},\tau 1}}}\right). \tag{34}$$

The magnitude $\dot{\theta}_{\mathrm{OG}}$ and the orientation $\theta_{\dot{\theta}_{\mathrm{OG}}}$ of the projected rotation rate on the LTP are then obtained as

$$\dot{\theta}_{\mathrm{OG}} = \sqrt{\dot{\theta}_{x_{\mathrm{LTP},\tau 1}}^2 + \dot{\theta}_{y_{\mathrm{LTP},\tau 1}}^2}, \text{ and} \tag{35}$$

$$\theta_{\dot{\theta}_{\mathrm{OG}}} = \operatorname{atan2}\left(\frac{\dot{\theta}_{y_{\mathrm{LTP},\tau 1}}}{\dot{\theta}_{x_{\mathrm{LTP},\tau 1}}}\right). \tag{36}$$

The orientation differences between $\theta_{\mathrm{yaw}_{\mathrm{LCP}}}$ and $\theta_{\dot{\theta}_{\mathrm{OG}}}$; and between $\theta_{\mathrm{yaw}_{\mathrm{LCP}}}$ and $\theta_{a_{\mathrm{OG}}}$ are obtained by

$$\Delta\left(\theta_{a_{\mathrm{OG}}}, \theta_{\mathrm{yaw}_{\mathrm{LCP}}}\right) = \theta_{a_{\mathrm{OG}}} - \theta_{\mathrm{yaw}_{\mathrm{LCP}}}, \text{ and} \tag{37}$$

$$\Delta\left(\theta_{\dot{\theta}_{\mathrm{OG}}}, \theta_{\mathrm{yaw}_{\mathrm{LCP}}}\right) = \theta_{\dot{\theta}_{\mathrm{OG}}} - \theta_{\mathrm{yaw}_{\mathrm{LCP}}}. \tag{38}$$

Finally, the acceleration and rotation rates are projected on the vehicle yaw as

$$\begin{bmatrix} a_{x_{\mathrm{LCP}}}^{\mathrm{LTP}} \\ a_{y_{\mathrm{LCP}}}^{\mathrm{LTP}} \end{bmatrix} = \begin{bmatrix} a_{\mathrm{OG}} \cos\left(\Delta\left(\theta_{a_{\mathrm{OG}}}, \theta_{\mathrm{yaw}_{\mathrm{LCP}}}\right)\right) \\ a_{\mathrm{OG}} \sin\left(\Delta\left(\theta_{a_{\mathrm{OG}}}, \theta_{\mathrm{yaw}_{\mathrm{LCP}}}\right)\right) \end{bmatrix}, \text{ and} \tag{39}$$

$$\begin{bmatrix} \dot{\theta}_{\mathrm{roll}_{\mathrm{LCP}}}^{\mathrm{LTP}} \\ \dot{\theta}_{\mathrm{pitch}_{\mathrm{LCP}}}^{\mathrm{LTP}} \end{bmatrix} = \begin{bmatrix} \dot{\theta}_{\mathrm{OG}} \cos\left(\Delta\left(\theta_{\dot{\theta}_{\mathrm{OG}}}, \theta_{\mathrm{yaw}_{\mathrm{LCP}}}\right)\right) \\ \dot{\theta}_{\mathrm{OG}} \sin\left(\Delta\left(\theta_{\dot{\theta}_{\mathrm{OG}}}, \theta_{\mathrm{yaw}_{\mathrm{LCP}}}\right)\right) \end{bmatrix}, \tag{40}$$

where $a_{x_{\mathrm{LCP}}}^{\mathrm{LTP}}$ and $a_{y_{\mathrm{LCP}}}^{\mathrm{LTP}}$ are the longitudinal and lateral accelerations, $\dot{\theta}_{\mathrm{roll}_{\mathrm{LCP}}}^{\mathrm{LTP}}$ and $\dot{\theta}_{\mathrm{pitch}_{\mathrm{LCP}}}^{\mathrm{LTP}}$ are the roll and pitch rates of the vehicle when projected on the ($x_{\mathrm{LTP}}$, $y_{\mathrm{LTP}}$)–plane. These quantities express the motion of the vehicle over the ($x_{\mathrm{LTP}}$, $y_{\mathrm{LTP}}$)–plane, and they are the inputs for the motion models used later. A graphical description of the projection of the IMU measurements can be seen in the Figure 4.

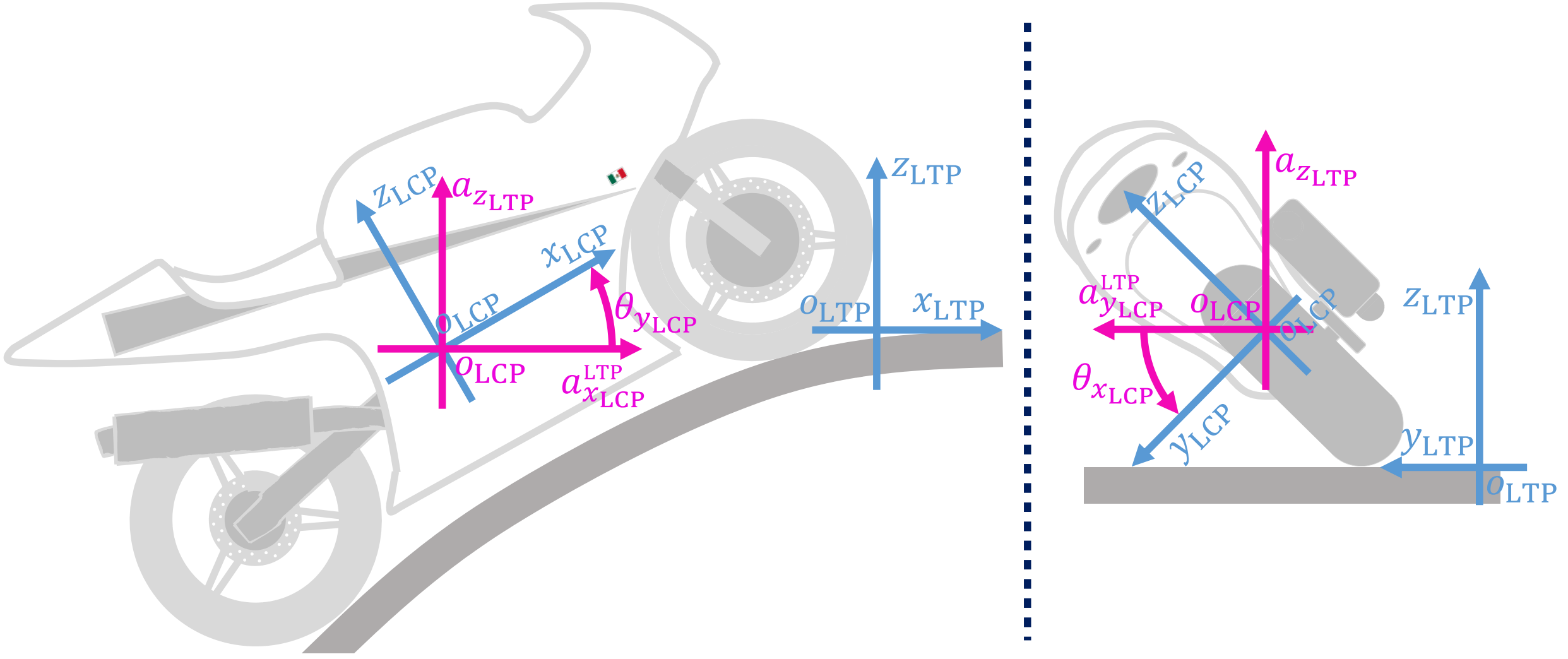

Figure 4. Left: a motorcycle driving up a parking lot ramp. Right: a motorcycle driving a left curve. With blue, the axes of the LCP and the LTP. With pink, the projection of the IMU measurements.

## 5. LiDAR-based Positioning Method

### *5.1. Introduction and Motivation*

As explained above, one of the most important positioning references, GPS-RTK, is not always available; and other positioning methods do not have the high accuracy requirements for automotive research. So, there is a necessity for better performing methods, which are practical as well. In this work, a novel, high accuracy and robust method is presented. This is explained in detail following.

### *5.2. Infrastructure markers*

The proposed positioning concept relies on the precise distance measurements of a mechanical LiDAR sensor [10] to infra-red reflecting infrastructure elements (markers). In this work, a high reflective tape is used [11]. This tape complies with the UN/ECE 104 norm [12], which regulates retro-reflectors for trucks and trailers in Europe. However, the markers can be, for example, traffic signs, which are common and already present in most parking houses. New markers can be easily added by looking at the blueprints of the building where they are placed. The effort of making new measurements is manageable as well, since devices as a Tachiometer simplify greatly this task. As explained later, the markers should be high reflective, as is common for traffic signs.

The number of markers in the interest area influences how often the vehicle state can be estimated. A small number of markers means more infrequent estimations, and a larger number allows more frequent estimations. It should be taken into account that if too many markers are present, this can lead to confusion in the marker identification. In the tests performed, the maximum distance at which a marker is seen with the desired intensity (more about this later), is $d_{\mathrm{velo,max}} \approx 16\mathrm{m}$. A marker library is generated beforehand, and saved for later processing.

### 5.3. *Point cloud managing*

The selected LiDAR is able to measure the National Institute for Standards and Technology (NIST) reflectivity [13]. Considering that a constraint for the algorithm is the runtime, it is the NIST-calibrated reflectivity the first filtering criteria for the point cloud. So, a reflectivity threshold of $\Gamma_{\iota,\mathrm{velo}} = 200$ is set. All points with less reflectivity are filtered out, since they most likely represent non-reflective infrastructure elements, such as walls, the roof or the floor. All points with higher reflectivity are kept, as they most likely correspond to markers.

For being able to process the LiDAR packets as they are generated, the points are clustered according to their timestamp. For this, a threshold of $\Gamma_{\tau_{\mathrm{c}},\mathrm{velo}} = 0.5\ \mathrm{ms}$ is set. The chosen mechanical LiDAR has a maximum rotation rate of $\omega_{\mathrm{velo,max}} = 1200$ rpm. So, the maximum horizontal Euclidean distance for point clustering is $d_{\mathrm{c,max}} = d_{\mathrm{velo,max}} \cdot \sin\left(\omega_{\mathrm{velo,max}} \cdot 6 \cdot \Gamma_{\tau_{\mathrm{c,velo}}}\right) = 57.89$ cm. This is the theoretical minimum distance between markers to avoid confusions. Once the points are assigned to a cluster, the properties of the cluster (position, timestamp) are calculated by using the mid-range arithmetic mean $\left(\frac{\max - \min}{2}\right)$ for the values of the points within the cluster.

### 5.4. *Marker identification*

To identify the markers, a rough position $\boldsymbol{p}_{\mathrm{r}} = [x_{\mathrm{r}}, y_{\mathrm{r}}]^{\mathrm{T}}$ and a rough orientation $\psi_{\mathrm{r}}$ estimates of the LCP are required. These can come from other *inaccurate* information sources or initial conditions for the test that are manually set. Assuming the true vehicle position is $\boldsymbol{p}_{\mathrm{t}} = [x_{\mathrm{t}}, y_{\mathrm{t}}]^{\mathrm{T}}$ and the true orientation is $\psi_{\mathrm{t}}$, then $\boldsymbol{p}_{\mathrm{r}}$ is constrained by $|\overrightarrow{\boldsymbol{p}_{\mathrm{r}}\,\boldsymbol{p}_{\mathrm{t}}}| < \frac{d_{\mathrm{c,max}}}{2}$; and $\psi_{\mathrm{r}}$ is constrained by $|\psi_{\mathrm{t}} - \psi_{\mathrm{r}}| < 0.78$ rad. By using $\psi_{\mathrm{r}}$, the LCP is oriented according to the LTP. Then, with a LiDAR measurement in LCP $\boldsymbol{p}_{\mathrm{m}} = [x_{\mathrm{m}}, y_{\mathrm{m}}]^{\mathrm{T}}$, the apparent position of the marker in LTP is given by $\boldsymbol{p}_{\mathrm{s}} = \boldsymbol{p}_{\mathrm{r}} + \boldsymbol{p}_{\mathrm{m}}$. This is compared to the marker library, and the closest marker to this position is assumed to be the seen one.

### 5.5. Velocity estimation

For the velocity calculation, it is assumed that the vehicle moves with constant velocity over an arch with constant radius between two consecutive measurements. Then, a cone shape is formed from two LiDAR measurements $\boldsymbol{p}_{\mathrm{m1}} = [x_{\mathrm{m1}}, y_{\mathrm{m1}}]^{\mathrm{T}}$ and $\boldsymbol{p}_{\mathrm{m2}} = [x_{\mathrm{m2}}, y_{\mathrm{m2}}]^{\mathrm{T}}$, done at time instances $\tau_1$ and $\tau_2$ respectively. So, let $d_{\mathrm{m1}}$, $d_{\mathrm{m2}}$, $\theta_{\mathrm{m1}}$ and $\theta_{\mathrm{m2}}$be the distance and azimuth measurements corresponding to $\boldsymbol{p}_{\mathrm{m1}}$ and $\boldsymbol{p}_{\mathrm{m2}}$accordingly. Also, given that the LCP moves relative to the LTP during $\Delta\tau = \tau_2 - \tau_1$, it is denoted as $(\mathrm{LCP}, \tau_1)$ and $(\mathrm{LCP}, \tau_2)$. With this information, the internal angles of the cone can then be calculated as

$$\theta_{\mathrm{v1}} = \begin{cases} \theta_{\mathrm{m1}}, & \theta_{\mathrm{m1}} \leq \pi \\ 2\pi - \theta_{\mathrm{m1}}, & \theta_{\mathrm{m1}} > \pi \end{cases}, \text{ and} \tag{41}$$

$$\theta_{\mathrm{v2}} = \begin{cases} \pi - \theta_{\mathrm{m2}} + \Delta\tau\dot{\theta}_{\mathrm{yaw_{LTP}}}, & \theta_{\mathrm{m2}} \leq \pi \\ \theta_{\mathrm{m2}} + \Delta\tau\dot{\theta}_{\mathrm{yaw_{LTP}}} - \pi, & \theta_{\mathrm{m2}} > \pi \end{cases}. \tag{42}$$

Then, the distance that the vehicle travels during $\Delta\tau$ can be approximated by the cosine law as

$$d_{\mathrm{car}} = \sqrt{d_{\mathrm{m1}}^2 + d_{\mathrm{m2}}^2 - 2d_{\mathrm{m1}}d_{\mathrm{m2}}\cos\left(\pi - \theta_{\mathrm{v1}} - \theta_{\mathrm{v2}} + \Delta\tau\dot{\theta}_{\mathrm{yaw_{LTP}}}\right)}. \tag{43}$$

Assuming the vehicle moves with constant velocity over ground $v_{o_{\mathrm{LCP}}}$ and constant $\dot{\theta}_{\mathrm{yaw_{LTP}}}$ during $\Delta\tau$, then

$$v_{o_{\mathrm{LCP}}} = \frac{d_{\mathrm{car}}}{\Delta\tau}. \tag{44}$$

A graphical description of this procedure can be seen in the Figure 5.

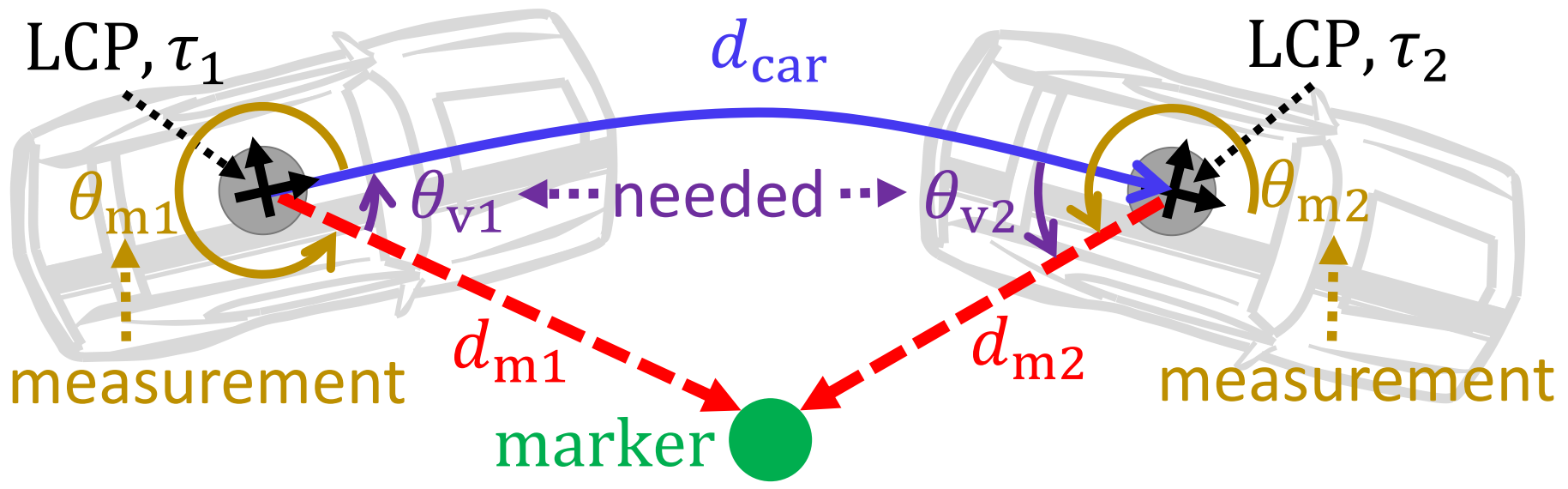

Figure 5. Graphical representation of the procedure for velocity calculation.

This procedure cannot be used for the pose estimation, as it is undefined around the seen marker.

### 5.6. Pose estimation

The orientation ambiguity mentioned in the previous section is solved when the measurements $\boldsymbol{p}_{\mathrm{m1}}$ and $\boldsymbol{p}_{\mathrm{m2}}$ correspond to different markers. With the same assumption that the vehicle moves with constant $v_{o_{\mathrm{LCP}}}$ and constant $\dot{\theta}_{\mathrm{yaw_{LTP}}}$ during $\Delta\tau$, then the radius of the arch the vehicles drives over is given by

$$r_{d_{\mathrm{car}}} = \frac{v_{o_{\mathrm{LCP}}}}{\dot{\theta}_{\mathrm{yaw_{LTP}}}}. \tag{45}$$

So, the movement of $o_{\mathrm{LCP}}$ on the $(x_{\mathrm{LTP}}, y_{\mathrm{LTP}})$–plane during $\Delta\tau$ expressed in $(\mathrm{LCP}, \tau_1)$ is given by

$$\begin{bmatrix} \Delta x_{o_{\mathrm{LCP}}}^{LTP} \\ \Delta y_{o_{\mathrm{LCP}}}^{LTP} \\ \Delta\theta_{\mathrm{yaw_{LCP}}} \end{bmatrix} = \begin{bmatrix} r_{d_{\mathrm{car}}} \sin\left(\Delta\tau \dot{\theta}_{\mathrm{yaw_{LTP}}}\right) \\ r_{d_{\mathrm{car}}} - \left(r_{d_{\mathrm{car}}} \cos\left(\Delta\tau \dot{\theta}_{\mathrm{yaw_{LTP}}}\right)\right) \\ \Delta\tau \dot{\theta}_{\mathrm{yaw_{LTP}}} \end{bmatrix}. \tag{46}$$

Given that $\Delta\theta_{\mathrm{yaw_{LCP}}}$ is known, then $\boldsymbol{p}_{\mathrm{m2}}$ can be rotated as if $(\mathrm{LCP}, \tau_2)$ had the same orientation as $(\mathrm{LCP}, \tau_1)$. Then, if $\boldsymbol{p}_{\mathrm{n1}} = [x_{\mathrm{n1}}, y_{\mathrm{n1}}]^{\mathrm{T}}$ and $\boldsymbol{p}_{\mathrm{n2}} = [x_{\mathrm{n2}}, y_{\mathrm{n2}}]^{\mathrm{T}}$ are the positions in LTP of the markers $\boldsymbol{p}_{\mathrm{n1}}$ and $\boldsymbol{p}_{\mathrm{n2}}$ seen at the measurements $\boldsymbol{p}_{\mathrm{m1}}$ and $\boldsymbol{p}_{\mathrm{m2}}$ respectively, then the position $\boldsymbol{p}_{\mathrm{n2}}$ in $(\mathrm{LCP}, \tau_1)$ is given by

$$\boldsymbol{p}_{\mathrm{n2}}^{\mathrm{LCP},\tau_1} = \boldsymbol{p}_{\mathrm{n2}} + \begin{bmatrix} \Delta x_{o_{\mathrm{LCP}}}^{LTP} & \Delta y_{o_{\mathrm{LCP}}}^{LTP} \end{bmatrix}^{\mathrm{T}} . \tag{47}$$

Now, assuming $x_{\mathrm{n1}} < x_{\mathrm{n2}}$, the orientation of $\overrightarrow{\boldsymbol{p}_{\mathrm{n1}}\boldsymbol{p}_{\mathrm{n2}}^{\mathrm{LCP},\tau_1}}$ in $(\mathrm{LCP}, \tau_1)$ is given by

$$\theta_{\mathrm{n1n2}}^{\mathrm{LCP},\tau_1} = \mathrm{arctan2}\left(y_{\mathrm{n2}}^{\mathrm{LCP},\tau_1} - y_{\mathrm{n1}}, x_{\mathrm{n2}}^{\mathrm{LCP},\tau_1} - x_{\mathrm{n1}}\right) , \tag{48}$$

and in LTP is expressed by

$$\theta_{\mathrm{n1n2}}^{\mathrm{LTP}} = \mathrm{arctan2}(y_{\mathrm{n2}} - y_{\mathrm{n1}}, x_{\mathrm{n2}} - x_{\mathrm{n1}}) . \tag{49}$$

Remembering the yaw definition, then the yaw angle of the vehicle $\theta_{\mathrm{yaw_{LCP}}}$ at time instances $\tau_1$ and $\tau_2$ is calculated as

$$\begin{bmatrix} \theta_{\mathrm{yaw_{LCP,\tau_1}}} \\ \theta_{\mathrm{yaw_{LCP,\tau_2}}} \end{bmatrix} = \begin{bmatrix} \theta_{\mathrm{n1n2}}^{\mathrm{LTP}} - \theta_{\mathrm{n1n2}}^{\mathrm{LCP},\tau_1} \\ \theta_{\mathrm{n1n2}}^{\mathrm{LTP}} - \theta_{\mathrm{n1n2}}^{\mathrm{LCP},\tau_1} + \Delta\theta_{\mathrm{yaw_{LCP}}} \end{bmatrix} . \tag{50}$$

Since $\boldsymbol{p}_{\mathrm{n1}}$ and $\boldsymbol{p}_{\mathrm{n2}}^{\mathrm{LCP},\tau_1}$ are now expressed in $(\mathrm{LCP}, \tau_1)$, they are rotated by $\theta_{\mathrm{yaw_{LCP,\tau_1}}}$, so

$$\begin{bmatrix} \boldsymbol{p}_{\mathrm{n1}}^{\theta_{\mathrm{yaw_{LCP,\tau_1}}}} \\ \boldsymbol{p}_{\mathrm{n2}}^{\mathrm{LCP},\tau_1,\theta_{\mathrm{yaw_{LCP,\tau_1}}}} \end{bmatrix} = \begin{bmatrix} \boldsymbol{R}\left(\theta_{\mathrm{yaw_{LCP,\tau_1}}}\right)\boldsymbol{p}_{\mathrm{n1}} \\ \boldsymbol{R}\left(\theta_{\mathrm{yaw_{LCP,\tau_1}}}\right)\boldsymbol{p}_{\mathrm{n2}}^{\mathrm{LCP},\tau_1} \end{bmatrix} , \tag{51}$$

where

$$\boldsymbol{R}\left(\theta_{\mathrm{yaw_{LCP,\tau_1}}}\right) = \begin{bmatrix} \cos\left(\theta_{\mathrm{yaw_{LCP,\tau_1}}}\right) & -\sin\left(\theta_{\mathrm{yaw_{LCP,\tau_1}}}\right) \\ \sin\left(\theta_{\mathrm{yaw_{LCP,\tau_1}}}\right) & \cos\left(\theta_{\mathrm{yaw_{LCP,\tau_1}}}\right) \end{bmatrix} . \tag{52}$$

Later, the linear offsets from $(\mathrm{LCP}, \tau_1)$ to the LTP are calculated as

$$\begin{bmatrix} x^{\mathrm{LTP}}_{o_{\mathrm{LCP}},\tau_1} \\ y^{\mathrm{LTP}}_{o_{\mathrm{LCP}},\tau_1} \end{bmatrix} = \frac{\left(\boldsymbol{p}_{\mathrm{n1}} - \boldsymbol{p}_{\mathrm{n1}}^{\theta \mathrm{yaw}_{\mathrm{LCP},\tau_1}}\right) + \left(\boldsymbol{p}_{\mathrm{n2}} - \boldsymbol{p}_{\mathrm{n2}}^{\mathrm{LCP},\tau_1,\theta \mathrm{yaw}_{\mathrm{LCP},\tau_1}}\right)}{2}. \tag{53}$$

By definition, $x^{\mathrm{LTP}}_{o_{\mathrm{LCP}},\tau_1}$ and $y^{\mathrm{LTP}}_{o_{\mathrm{LCP}},\tau_1}$ represent the (x,y) coordinates in LTP of $o_{\mathrm{LCP}}$ at $\tau_1$. Finally, knowing $\Delta x^{\mathrm{LTP}}_{o_{\mathrm{LCP}}}$ and $\Delta y^{\mathrm{LTP}}_{o_{\mathrm{LCP}}}$, the linear offsets from $(\mathrm{LCP}, \tau_2)$ to the LTP are calculated as

$$\begin{bmatrix} x^{\mathrm{LTP}}_{o_{\mathrm{LCP}},\tau_2} \\ y^{\mathrm{LTP}}_{o_{\mathrm{LCP}},\tau_2} \end{bmatrix} = \begin{bmatrix} x^{\mathrm{LTP}}_{o_{\mathrm{LCP}},\tau_1} \\ y^{\mathrm{LTP}}_{o_{\mathrm{LCP}},\tau_1} \end{bmatrix} + \begin{bmatrix} \Delta x^{\mathrm{LTP}}_{o_{\mathrm{LCP}}} \\ \Delta y^{\mathrm{LTP}}_{o_{\mathrm{LCP}}} \end{bmatrix}. \tag{54}$$

By definition, $x^{\mathrm{LTP}}_{o_{\mathrm{LCP}},\tau_2}$ and $y^{\mathrm{LTP}}_{o_{\mathrm{LCP}},\tau_2}$ represent the (x,y) coordinates in LTP of $o_{\mathrm{LCP}}$ at $\tau_2$. A graphical description of this procedure is seen in the Figure 6.

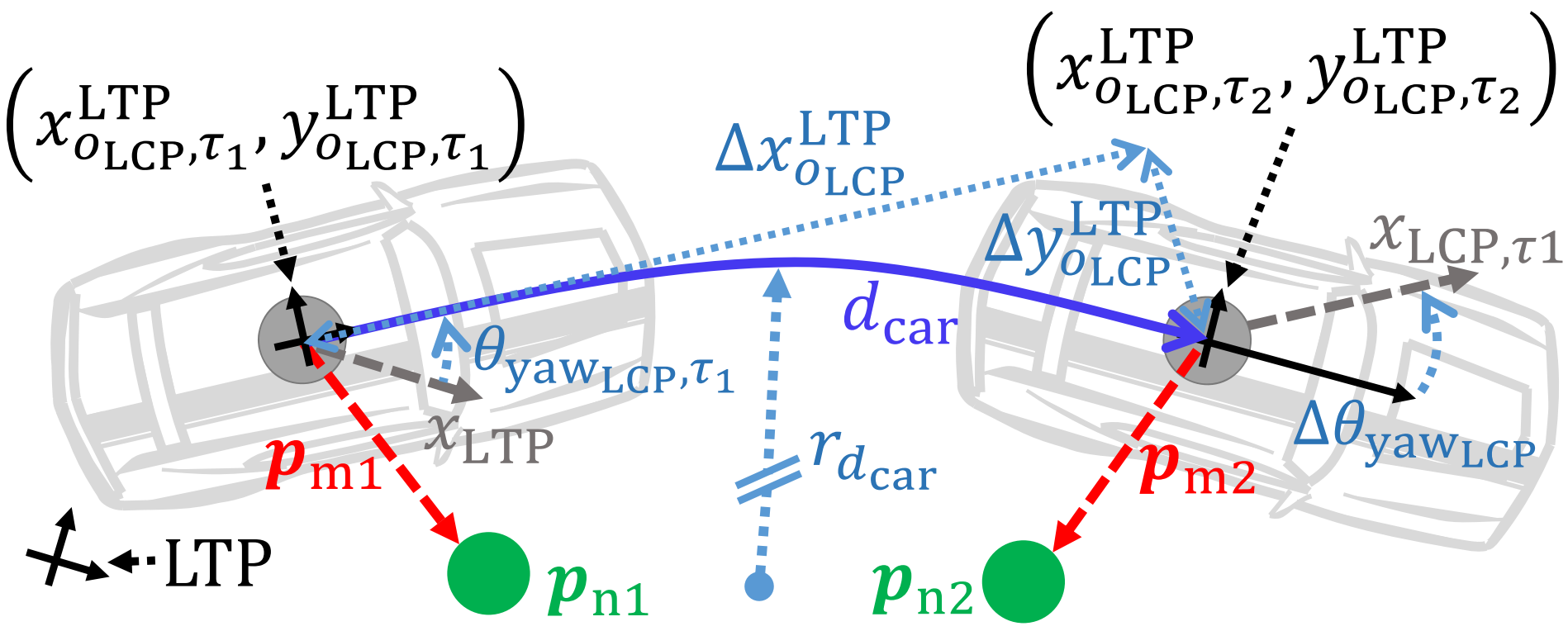

Figure 6. Graphical representation of the procedure of pose calculation.

## 6. Evaluation and results

For an adequate evaluation, the proposed method is first installed on an open-air test track for the sole purpose of using an ADMA-G-Pro+ as reference. The setup is later installed in a closed test hall for validating indoor use possibilities. Both evaluations are described in detail in the following.

### *6.1. Outdoor evaluation*

As stated above, a widely accepted reference for vehicle state estimation is the combination of IMUs and GPS with RTK correction data. Because of this, an ADMA G-Pro+ is used as benchmark for evaluating the proposed LbPM method.

First, the markers are put on fixed poles around the test track, and their positions are measured with a GPS antenna with RTK correction data, so a position accuracy of at least 2 cm is obtained. The marker library with identifiers and positions is saved.

So as to be able to identify possible weaknesses of the proposed method, two different maneuvers where driven: a drive-by along the markers and a slalom along one side of the markers. Both maneuvers are driven with different velocities as well, ranging from 5 km/h and up to 40 km/h. The results can be seen on the Tables 1, 2 and 3. It is important to note that the presented method receives no correction data from the SatNav system.

Table 1: Velocity accuracy results for the proposed LbPM. Shown are the driven maneuvers, the velocity at which the maneuvers were driven, the mean deviation from the reference, standard deviation of the deviation from the reference and maximum deviation from the reference.

| **Drive-by at** | **Mean ($m/s$)** | **Std. dev. ($m/s$)** | **Max ($m/s$)** |
|---|---|---|---|
| 5 km/h | 0.06 | 0.08 | 0.33 |
| 10 km/h | 0.08 | 0.10 | 0.57 |
| 15 km/h | 0.07 | 0.09 | 0.39 |
| 20 km/h | 0.08 | 0.09 | 0.50 |
| 25 km/h | 0.08 | 0.10 | 0.65 |
| 30 km/h | 0.08 | 0.10 | 0.57 |
| 35 km/h | 0.08 | 0.11 | 0.44 |
| 40 km/h | 0.11 | 0.13 | 0.47 |
| **Slalom at** | **Mean ($m/s$)** | **Std. dev. ($m/s$)** | **Max ($m/s$)** |
| 5 km/h | 0.08 | 0.11 | 0.59 |
| 10 km/h | 0.09 | 0.12 | 0.59 |
| 20 km/h | 0.14 | 0.17 | 0.71 |
| 30 km/h | 0.18 | 0.24 | 0.76 |
| 40 km/h | 0.18 | 0.22 | 0.62 |

Table 2: Position accuracy results for the proposed LbPM. Shown are the driven maneuvers, the velocity at which the maneuvers were driven, the mean deviation from the reference, standard deviation of the deviation from the reference and maximum deviation from the reference.

| **Drive-by at** | **Mean ($m$)** | **Std. dev. ($m$)** | **Max ($m$)** |
|---|---|---|---|
| 5 km/h | 0.04 | 0.02 | 0.09 |
| 10 km/h | 0.03 | 0.02 | 0.10 |
| 15 km/h | 0.03 | 0.02 | 0.13 |
| 20 km/h | 0.03 | 0.02 | 0.09 |
| 25 km/h | 0.04 | 0.02 | 0.07 |
| 30 km/h | 0.06 | 0.02 | 0.10 |
| 35 km/h | 0.07 | 0.03 | 0.11 |
| 40 km/h | 0.08 | 0.03 | 0.15 |
| **Slalom at** | **Mean ($m$)** | **Std. dev. ($m$)** | **Max ($m$)** |
| 5 km/h | 0.04 | 0.02 | 0.12 |
| 10 km/h | 0.04 | 0.02 | 0.13 |
| 20 km/h | 0.04 | 0.02 | 0.10 |
| 30 km/h | 0.05 | 0.02 | 0.09 |
| 40 km/h | 0.10 | 0.02 | 0.12 |

Table 3: Orientation accuracy results for the proposed LbPM. Shown are the driven maneuvers, the velocity at which the maneuvers were driven, the mean deviation from the reference, standard deviation of the deviation from the reference and maximum deviation from the reference.

| Drive-by at | Mean (deg) | Std. dev. (deg) | Max (deg) |
|---|---|---|---|
| 5 km/h | 0.73 | 0.25 | 1.48 |
| 10 km/h | 0.19 | 0.20 | 0.86 |
| 15 km/h | 0.26 | 0.19 | 0.69 |
| 20 km/h | 0.37 | 0.23 | 0.83 |
| 25 km/h | 0.58 | 0.23 | 0.84 |
| 30 km/h | 0.51 | 0.22 | 0.96 |
| 35 km/h | 0.44 | 0.25 | 0.88 |
| 40 km/h | 0.41 | 0.26 | 0.86 |
| **Slalom at** | **Mean (deg)** | **Std. dev. (deg)** | **Max (deg)** |
| 5 km/h | 0.24 | 0.29 | 1.37 |
| 10 km/h | 0.40 | 0.29 | 1.22 |
| 20 km/h | 0.32 | 0.36 | 1.18 |
| 30 km/h | 0.36 | 0.40 | 1.28 |
| 40 km/h | 0.53 | 0.43 | 1.25 |

### *6.2. Indoor evaluation*

After the evaluation outdoors, the setup is installed in a closed test hall. Given that there is no positioning system with a better performance than the one presented, it is not useful to generate comparison numbers for the indoor tests. Because of this, the evaluation is limited to the system stability and the visual confirmation that the position obtained from the LbPM does describe the trajectory that was driven with the test vehicle.

For the indoor evaluation, a trajectory is driven in a closed hall. The trajectory includes straight sections as well as a slalom. The trajectory is driven with various velocities, going from 5 km/h and up to 30 km/h. Due to space limitations in the test hall, tests with higher velocities are not performed. The Figure 7 shows a test at 10 km/h. As can be seen, the position generated by the LbPM is stable during the whole test. This is recognized from the absence of sharp corners that do not describe how a road vehicle moves. The soft, constant curves obtained as output from the LbPM do depict how the vehicle is driven.

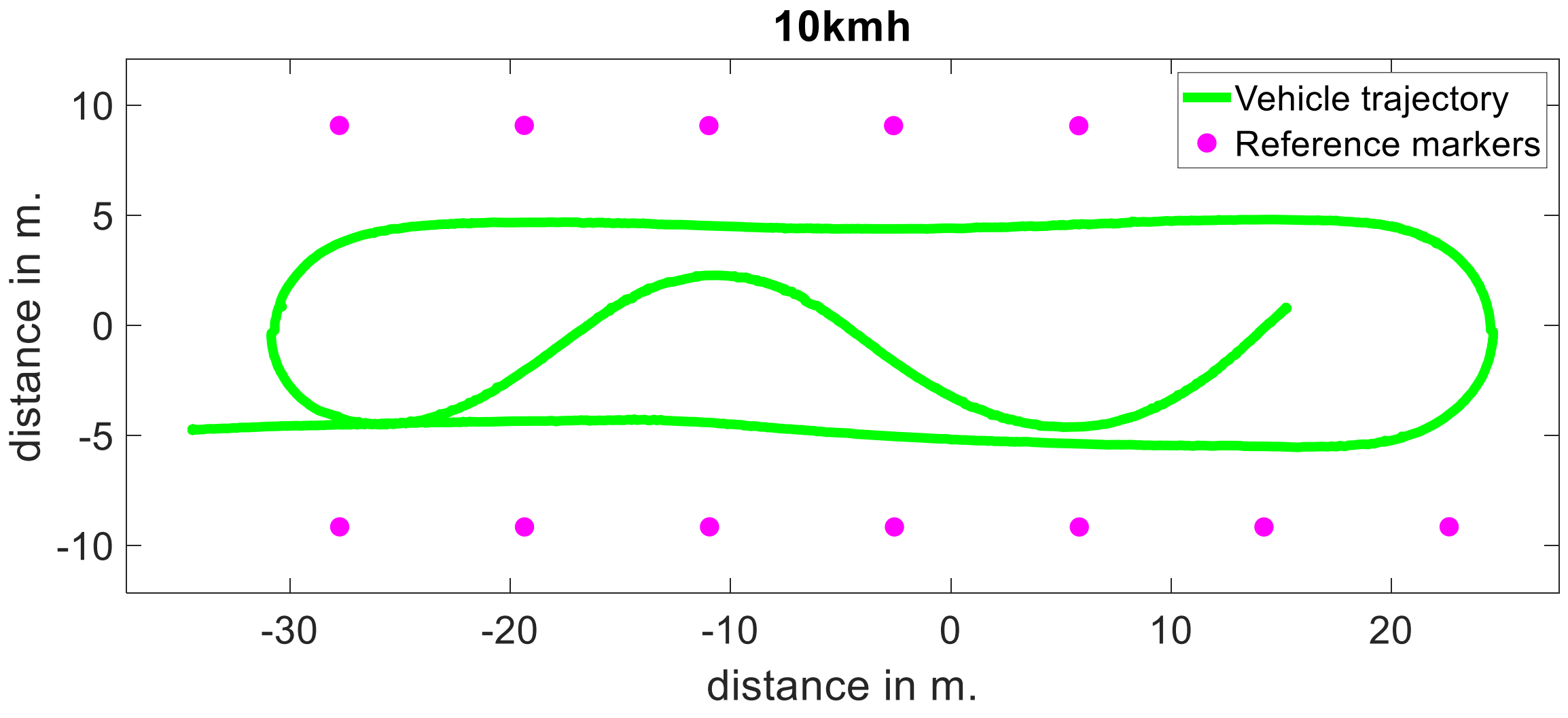

Figure 7. Position output from the proposed LbPM method. With pink, the reference markers. With green, the trajectory driven by the test vehicle.

### *6.3. Runtime evaluation*

An important constraint for algorithms meant for vehicle safety, is the runtime. For evaluating the possibility of real-time implementation of the presented LbPM, runtime measurements are performed on an Intel i7-6820HQ CPU by using the Matlab profiler. The runtime is evaluated for the same algorithm as Mex-file, as well as high-level Matlab code. After measuring the runtime of 141,600 algorithm executions, a median runtime of 40.77 µs is obtained. This runtime includes velocity, orientation and position estimation. Considering the update rate of the chosen LiDAR of 20 Hz, the proposed method has the potential of real-time implementation. The detailed results can be seen in the Table 4.

Table 4: Runtime of the proposed LbPM when measured with the Matlab profiler.

| Estimation | Type | Median ($\mu s$) | Std. dev. ($\mu s$) | Max ($\mu s$) |
|---|---|---|---|---|
| Velocity | Matlab | 11.32 | 14.15 | 73.11 |
| Velocity | Mex | 20.76 | 27.99 | 117.45 |
| Pose | Matlab | 42.66 | 54.29 | 190.17 |
| Pose | Mex | 20.01 | 45.34 | 123.34 |

## 7. Conclusion and future work

In this work, a LiDAR-based Positioning Method is presented. The method includes the detailed procedure for projecting IMU measurements on the LTP, as well as the calculation of position, orientation and velocity for the vehicle.

The method is extensively tested by performing driving maneuvers with various velocities, and evaluated using an ADMA-G-Pro+, which confirms the high accuracy of the proposed method. This high accuracy, together with the stability that the system shows in the indoor tests, and the possibility of real-time implementation imply a massive progress for the indoor navigation.

The future work includes the implementation of the presented method in embedded hardware for on-line vehicle state estimation. It is planned as well to repeat the tests using different settings for the chosen LiDAR. This because a faster LiDAR rotation would imply faster updates, but a reduced accuracy.

## 8. Acknowledgements

The authors acknowledge the financial support by the Federal Ministry of Education and Research of Germany (BMBF) in the framework of FH-Impuls (project number 03FH7I02IA).